\documentclass[prd,twocolumn,reprint,preprintnumbers,nofootinbib,superscriptaddress]{revtex4-1}
\pdfoutput=1
\usepackage{cancel}

\input{header}

\begin{document}
\preprint{FERMILAB-PUB-21-588-T, CERN-TH-2021-186, DESY 21-188}
\title{Probing Neutrino-Portal Dark Matter at the Forward Physics Facility}

\author{Kevin J. Kelly}
\email{kjkelly@cern.ch}
\affiliation{Theoretical Physics Department, Fermilab, P.O. Box 500, Batavia, IL 60510, USA}
\affiliation{Theoretical Physics Department, CERN, Esplande des Particules, 1211 Geneva 23, Switzerland}

\author{Felix Kling}
\email{felix.kling@desy.de}
\affiliation{Theory Group, SLAC National Accelerator Laboratory, Menlo Park, CA 94025, USA}
\affiliation{Deutsches Elektronen-Synchrotron DESY, Notkestrasse 85, 22607 Hamburg, Germany}

\author{Douglas Tuckler}
\email{dtuckler@physics.carleton.ca}
\affiliation{Department of Physics, Carleton University, Ottawa, ON K1S 5B6, Canada}

\author{Yue Zhang}
\email{yzhang@physics.carleton.ca}
\affiliation{Department of Physics, Carleton University, Ottawa, ON K1S 5B6, Canada}

\begin{abstract}
The Forward Physics Facility (FPF), planned to operate near the ATLAS interaction point at the LHC, offers exciting new terrain to explore neutrino properties at TeV energy scales. It will reach an unprecedented regime for terrestrial neutrino experiments and provide the opportunity to reveal new physics of neutrinos at higher energy scales. We demonstrate that future detectors at the FPF have the potential to discover new mediators that couple predominantly to neutrinos, with masses between 0.3 and 20 GeV and small couplings not yet probed by existing searches. Such a neutrinophilic mediator is well motivated for addressing the origin of several neutrino-portal dark matter candidates, including thermal freeze-out and sterile-neutrino dark matter scenarios. Experimentally, the corresponding signatures include neutrino charged-current scattering events associated with large missing transverse momentum, and excessive apparent tau-neutrino events. We discuss the FPF detector capabilities needed for this search, most importantly the hadronic energy resolution.
\end{abstract}

\maketitle

\section{Introduction}\label{sec:Introduction}
While our experimental techniques of observing and understanding neutrinos is progressing rapidly in many different approaches, one particularly exciting opportunity is to observe interactions of neutrinos with very high energies in Earth-based detectors. Notable experiments in this direction include the IceCube Neutrino Observatory, that has been detecting atmospheric and astrophysical neutrinos of various energies~\cite{IceCube:2013cdw,LIGOScientific:2017ync}, as well as the recent first detection of collider-produced neutrinos by the FASER$\nu$ prototype at the Large Hadron Collider (LHC)~\cite{Abreu:2021hol}.
These experiments also open up a wide array of avenues for exploring fundamental properties of neutrinos in the context of beyond Standard Model (SM) physics, such as non-standard neutrino interactions~\cite{IceCubeCollaboration:2021euf}, new states in the neutrino sector~\cite{FASER:2019dxq,IceCube:2020phf}, and even tests of Lorentz-invariance violation~\cite{IceCube:2017qyp}.

New forces that act predominately on neutrinos are well-motivated candidates of new physics. They are often predicted in gauge or scalar extensions of the SM related to the mechanism for generating light neutrino masses, e.g., as in Refs.~\cite{Minkowski:1977sc,GellMann:1980vs,Yanagida:1979as,Glashow:1979nm,Mohapatra:1979ia,Schechter:1980gr,Foot:1988aq}. There are also strong motivations from the cosmic frontier that favor new neutrino interactions for explaining the origin of dark matter (DM) that fills the universe.
It is worth stressing that such an interaction is allowed to be much stronger than predicted in the SM~\cite{Ng:2014pca,Blinov:2019gcj}. Indeed, the LEP measurement of neutrino-$Z$ coupling through the $Z$ boson invisible width is only indirect and does not preclude neutrinos from having much stronger self interactions via the exchange of a new force carrier.
Strong neutrino self interactions are testable and can lead to a number of novel phenomena at neutrino and collider experiments~\cite{Lessa:2007up,Laha:2013xua,Brune:2018sab,Berryman:2018ogk,Kelly:2019wow,deGouvea:2019qaz,Brdar:2020nbj,Deppisch:2020sqh} that hunt for new states and make precision measurements, as well as in the early universe by leaving an imprint on Big Bang nucleosynthesis~\cite{Venzor:2020ova,Brinckmann:2020bcn} and the formation of the Cosmic Microwave Background~\cite{Kreisch:2019yzn,He:2020zns,Das:2020xke}.

We consider the scalar boson incarnation of the neutrinophilic force carrier. Such a scalar can be produced experimentally along with neutrinos or through bremsstrahlung off a neutrino beam. Its subsequent decay into to a neutrino and antineutrino pair usually appears invisible to detectors. However, it is possible to infer the occurrence of such a process by exploring the visible parts of the final state. We consider the neutrino beamstrahlung happening along with a charged-current (CC) interaction inside a fixed-target neutrino detector. The resulting final state is similar to that of a standard CC event but has novel features including charged lepton production carrying opposite lepton number (wrong-sign) to the incoming neutrino and sizable missing transverse momentum (MET) with respect to the incoming neutrino beam direction. To our knowledge, such a process was conceived first within the SM~\cite{Bardin:1970wq}, and later in models with a light Majoron~\cite{Barger:1981vd}. It is worth noting that these pioneering works focus on the wrong-sign charged lepton signature which would be a very clean signal in a detector with charge identification and if a neutrino beam is free from antineutrino pollution. If these criteria cannot be met, one needs to resort to MET as the key signal for the neutrinophilic scalar~\cite{Berryman:2018ogk}. In Ref.~\cite{Kelly:2019wow}, this was dubbed as the ``mono-neutrino'' signal, in analogy to the mono-$X$ searches widely performed at various colliders~\cite{Chekanov:2668732} to probe WIMP-like DM.

\begin{figure}[thbp]
\centering
\includegraphics[width=0.45\textwidth]{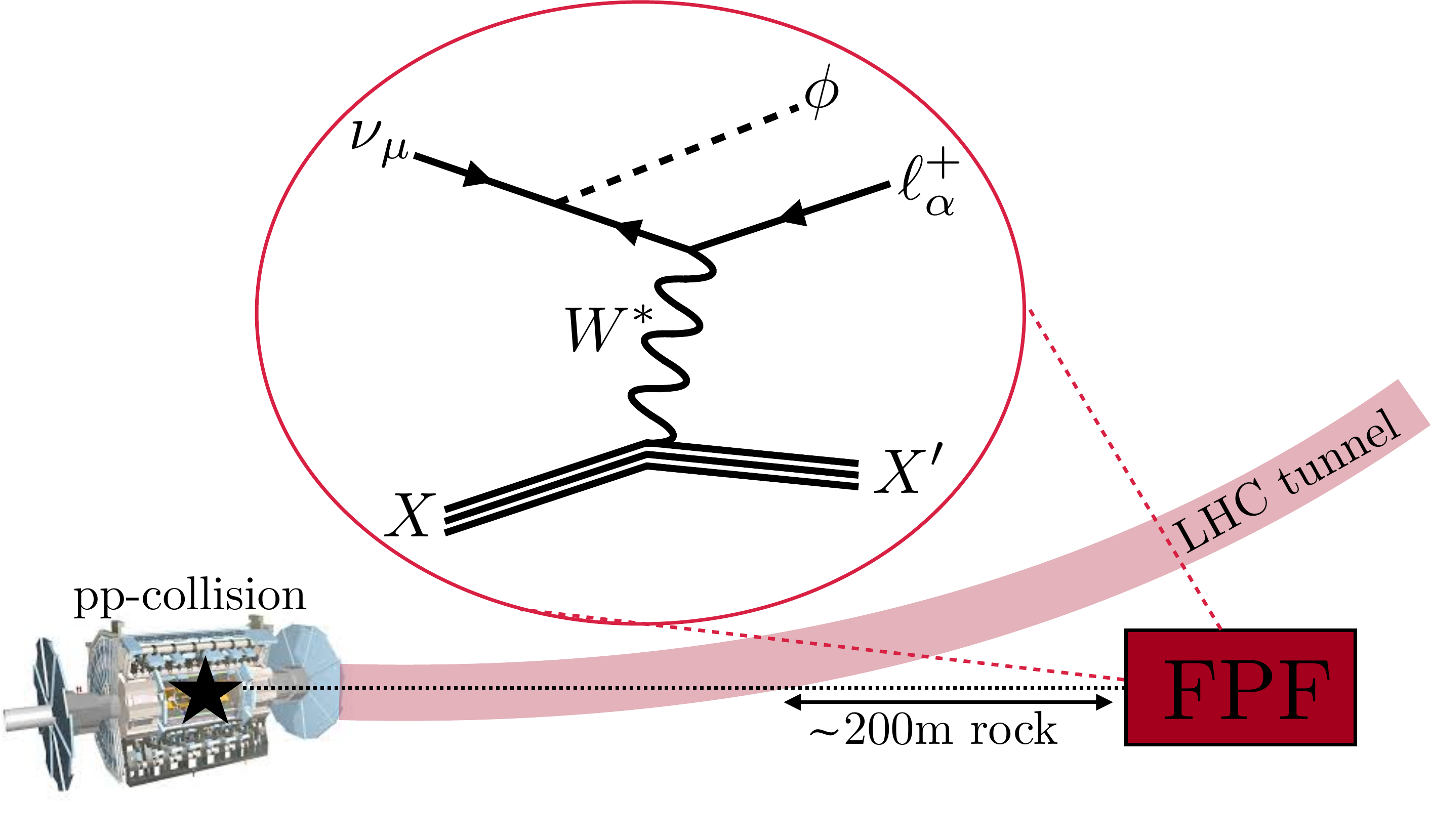}
\caption{Schematic representation of neutrino production (at the ATLAS interaction point) and scattering with a mono-neutrino signature in the FPF detector(s). The Feynman diagram (inset) depicts the signal process of interest for this work.}\label{fig:Feynman}
\end{figure}

In this article, we explore the potential of discovering a neutrinophilic scalar in the mono-neutrino channel using neutrinos originating from LHC proton-proton collisions and detectors located at the Forward Physics Facility (FPF) near the LHC interaction point~\cite{Anchordoqui:2021ghd}. A schematic picture is shown in \cref{fig:Feynman}. The target theory parameter space is motivated by previous works exploring the connection between neutrino self-interactions and the origin of DM~\cite{Kelly:2019wow, deGouvea:2019phk}. Clearly, the FPF setup has the advantage of searching for heavier neutrinophilic scalars compared to traditional accelerator neutrino experiments, thanks to the much higher typical energy (hundreds of GeV up to a few TeV) of neutrinos coming from the LHC. Moreover, the scattering of these high-energy neutrinos with the detector is deeply inelastic and well-described by the parton picture, resulting in much smaller nuclear uncertainty compared to the case of a GeV-scale neutrino beam. The corresponding neutrino fluxes are also better modeled than those at IceCube with a cosmogenic origin. 

This article is organized as follows.
\cref{sec:ModelDM} details the connection between neutrino self-interactions and the origin of DM relic density in several concrete models, and presents well-motivated targets for the FPF experiments to hopefully probe. In \cref{sec:Simulation}, we provide information about the simulations  performed for these scenarios. We attempt to keep our discussions detached from specific detector concepts, but we will address how various planned detectors (e.g., future upgrades to FASER$\nu$~\cite{FASER:2019dxq}, and the liquid argon FLArE proposal~\cite{Batell:2021blf}) connect to our results throughout. We perform two analyses -- \cref{sec:AnalysisMuon} discusses results for a neutrinophilic mediator with a flavor-diagonal coupling to muon-type neutrinos, whereas \cref{sec:AnalysisTau} allows for flavor off-diagonal couplings between the mediator, muon-type neutrinos, and tau-type neutrinos. Finally, \cref{sec:Conclusions} offers some concluding remarks.

\section{Neutrino Self-interaction Motivated by Dark Matter Origin} \label{sec:ModelDM}

As the benchmark model of our analysis, we introduce a massive scalar $\phi$ with the following low-energy coupling to neutrinos,
\begin{equation}\label{eq:Lagrangian}
    \mathcal{L} \supset \frac{1}{2} \lambda_{\alpha\beta} \nu_\alpha \nu_\beta \phi + \mathrm{h.c.} \ ,
\end{equation}
where $\alpha,\beta = e, \mu, \tau$ are flavor indices. Such an operator is not gauge invariant under the SM $SU(2)_L$ but could arise from a dimension-six or higher operator.
This benchmark model is the reminiscent of beyond SM ultraviolet completions where $\phi$ serves as, e.g., a lepton number charged scalar~\cite{Berryman:2018ogk}, or the Majoron~\cite{Gelmini:1980re,Chikashige:1980qk}.

The presence of the neutrinophilic force mediated by $\phi$ can also be used to address the origin of DM in our universe. 
Specifically, we will consider two classes of models where the DM relic density is produced either via thermal freeze out~\cite{Kelly:2019wow}, or a non-thermal freeze in mechanism~\cite{deGouvea:2019phk,Kelly:2020aks}.
In both cases, the interactions of $\phi$ introduced in~\cref{eq:Lagrangian} play an indispensable role in the origin of dark matter, which in turn provides well-motivated and highly-predictive targets for experimental tests.

\subsection{Thermal Freeze-Out Dark Matter}\label{subsec:thermalDM}

In this class of models, we consider $\phi$ serving as the portal between the visible and dark sectors.
In the early universe, DM thermalizes with SM neutrinos through the $\phi$ exchange. Its relic density is set by thermal freeze-out and annihilation into neutrinos.
We consider two possibilities, where dark matter is a Dirac fermion (DF) or complex scalar (CS), stabilized by a $Z_2$ or $Z_3$ symmetry, respectively.
The interaction Lagrangians are
\begin{equation}
\begin{split}
    \mathcal{L}_{\rm DF} &= \frac{1}{2}y \overline{\chi}^c \chi \phi + \mathrm{h.c.} \ , \\
    \mathcal{L}_{\rm CS} &= \frac{1}{6} y \chi^3 \phi + \mathrm{h.c.} \ .
    \end{split}
\end{equation}
In both cases, the operators are marginal thus $y$ is a dimensionless coupling. 
The dark matter stabilizing symmetries can be promoted to a $U(1)$ lepton number global symmetry, where $\phi$ carries charge\footnote{This is set by \cref{eq:Lagrangian}} $-2$ and $\chi$ carries charge $+1$ ($+2/3$) in the fermion (scalar) case.

The relevant annihilation cross sections for freeze out calculations in the two models are
\begin{equation}
\begin{split}
    \sigma v (\chi\chi \to \nu\nu) &= \frac{|\lambda_{\alpha\beta} y|^2 m_\chi^2 v^2}{16\pi (4m_\chi^2 - m_\phi^2)^2 (1+\delta_{\alpha \beta})} \ , \\
    \sigma v (\chi\chi \to \chi^* \nu\nu) &=  \frac{|\lambda_{\alpha\beta} y|^2}{2048 \pi^3 m_\chi^2 (1+\delta_{\alpha\beta})} \\
   &\hspace{0.4cm} \times \int_0^1 dz \frac{z\sqrt{(1-z)(9-z)}}{(m_\phi^2/m_\chi^2 -z)^2} \ ,
\end{split}
\end{equation}
where we assume only one element of the coupling matrix $\lambda_{\alpha\beta}$ is nonzero in each case.
We restrict our analysis to the heavy mediator case, with $m_\phi > m_\chi$, thus the relic density depends on both the DM and neutrino couplings to $\phi$.
For simplicity, we will not consider the case where $m_\phi < m_\chi$ and the $\chi\bar\chi\to \phi \phi^*$ annihilation channel also becomes relevant.
The relic density of dark matter in the two models has been computed in Ref.~\cite{Kelly:2019wow}.

\subsection{Sterile Neutrino Dark Matter} 

Another DM candidate we consider is sterile neutrino dark matter (S$\nu$DM), which is a mixture of a SM gauge singlet fermion $\nu_s$ and an active neutrino $\nu_a$,
\begin{equation}
    \nu_4 =  \nu_s\cos\theta + \nu_a \sin\theta \ ,
\end{equation}
where we denote the vacuum mixing angle as $\theta$. In the absence of $\phi$-mediated interactions, the production of S$\nu$DM was first explored by Dodelson and Widrow (DW)~\cite{Dodelson:1993je}, which occurs through active-sterile neutrino oscillation in the early universe, along with frequent weak interactions prior to neutrino decoupling. The resulting dark matter relic density is controlled by two parameters, $\theta$ and the dark matter mass $m_4$.
Such a simple and elegant mechanism has already been excluded by indirect detection searches for dark matter decaying into X-rays~\cite{Abazajian:2017tcc,Horiuchi:2013noa,Malyshev:2014xqa,Dessert:2018qih}.

Recently, Refs.~\cite{deGouvea:2019phk,Kelly:2020aks,Kelly:2020pcy} pointed out that new neutrino self interaction such as \cref{eq:Lagrangian} allows neutrinos to stay in thermal equilibrium with themselves for a longer period in early universe, thus enabling more efficient S$\nu$DM production with smaller mixing angles and opening up viable parameter space of S$\nu$DM in the keV $\lesssim m_4 \lesssim$ MeV mass window. 
More concretely, the phase space number density of S$\nu$DM can be computed in a similar fashion as the DW case
\begin{equation}
    f_{\nu_4} (E, T) = \int_0^\infty \frac{\Gamma  f_{\nu_a} dz}{4 H z} \sin^2\theta_{\rm eff} \ ,
\end{equation} 
where $z\sim 1/T$ encodes the passage of time and $H$ is the Hubble parameter. Here, $\Gamma$ is the sum of neutrino self interaction and weak interaction rates and $\theta_{\rm eff}$ is the in-medium active-sterile neutrino mixing angle, which depends on $z$ through $\Gamma$ and the high-temperature potential for the sterile neutrino. We refer to~\cite{deGouvea:2019phk} for more details of the relic density calculation. In contrast to thermal freeze-out scenarios discussed above in subsection~\ref{subsec:thermalDM}, here the building up of dark matter relic abundance can be considered as a ``freeze-in'' mechanism~\cite{Hall:2009bx} but with a time-varying, mixing-induced coupling parameter for the $S\nu$DM.

\subsection{Mediator Constraints \& Phenomenology}

The neutrino self-interaction parameter space relevant for the above freeze-out and freeze-in production mechanisms of dark matter are clearly well-motivated targets for experimental probes.
Indeed, the benchmark models for neutrino self interaction have received extensive exploration recently. A number of constraints and projections have been derived,
including from cosmology~\cite{Kelly:2020aks,Blinov:2019gcj}, astrophysics~\cite{Ng:2014pca,Esteban:2021tub}, precision measurements of meson and charged-lepton decays~\cite{Pasquini:2015fjv,Berryman:2018ogk}, high-energy collider measurements of the $Z$ and Higgs bosons~\cite{Berryman:2018ogk,Brdar:2020nbj},
as well as accelerator neutrino experiments~\cite{Kelly:2019wow}. In this work, we will compare our new projections against these various constraints.

In the upcoming section, we investigate the exciting potential of using the FPF to probe uncovered neutrino self interaction theory space tied to the above dark matter targets.
The FPF sees a large flux of high-energy neutrinos produced from collisions and dumps of the LHC beam.
Among all the neutrino species, it has been found that $\nu_\mu$ and $\overline{\nu}_\mu$ constitute the highest flux~\cite{Abreu:2019yak}.
Therefore, we will consider a neutrinophilic mediator $\phi$ with coupling to at least one muon neutrino, $\lambda_{\mu\beta}$. For the second flavor index, we consider $\beta = \mu, \tau$, because they lead to muon or tau lepton production which have characteristic signatures in neutrino detectors.
The processes we explore involve the radiation of $\phi$ off the incoming neutrino beam when a charged current interaction takes place, as depicted in \cref{fig:Feynman}, followed by invisible decay of $\phi$ into neutrinos or dark matter.
In particular, we will explore the missing transverse momentum as the signal of a flavor-diagonal $\lambda_{\mu\mu}$ coupling,
and excessive $\tau$ lepton appearance as the signal of a flavor-off-diagonal $\lambda_{\mu\tau}$ coupling.

\section{Detector and Simulation}\label{sec:Simulation}

\begin{figure}
    \centering
    \includegraphics[width=\linewidth]{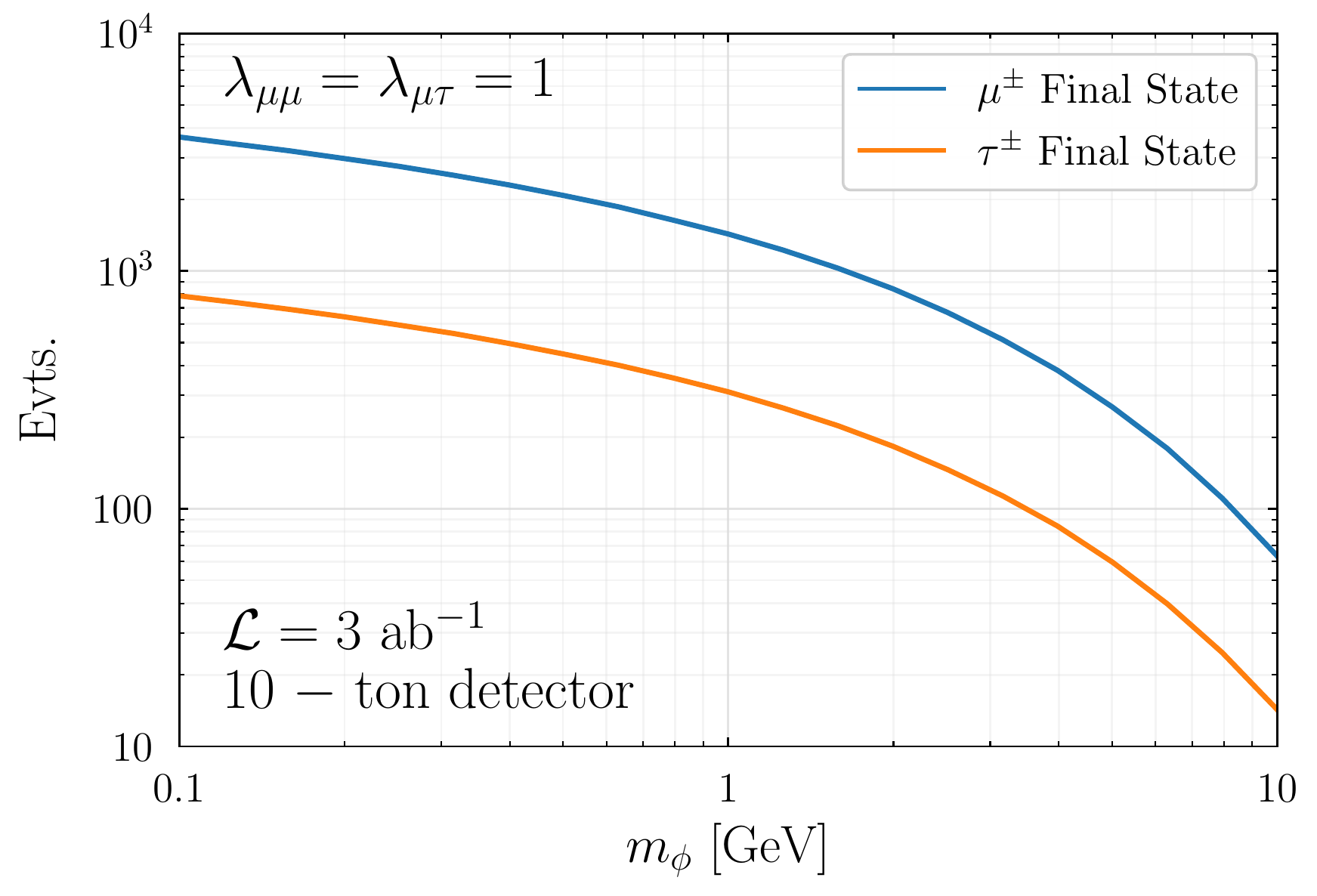}
    \caption{Total signal event rate for different scalar masses $m_\phi$ assuming either $\lambda_{\mu\mu}$ = 1 (blue) or $\lambda_{\mu\tau} = 1$ (orange). We have assumed an LHC luminosity of $3$ ab$^{-1}$ with a detector mass of 10 tons in the FPF.\label{fig:TotalRate}}
\end{figure}

The LHC is not only the particle collider with the highest collision energy, but also the source of the highest energy neutrinos made by humankind. These neutrinos are primarily produced by the decay of hadrons and form a strongly collimated beam of TeV energy neutrinos pointing in the forward direction. Proposals to utilize this neutrino beam date back to the early proposal stages of the LHC~\cite{DeRujula:1984pg, Vannucci:253670, DeRujula:1992sn, Park:2011gh, Buontempo:2018gta, Beni:2019gxv, XSEN:2019bel, Beni:2020yfy}. The situation changed in 2018, when the FASER collaboration placed a prototype neutrino detector employing emulsion films in the far-forward region of the ATLAS experiment. Despite the small target mass of about 12~kg and the small exposure time of a few weeks, the analysis reports the observation of six neutrino interaction candidates~\cite{Abreu:2021hol}, indicating the potential of the far-forward direction for neutrino measurements.

Starting in 2022, two dedicated neutrino detectors, FASER$\nu$~\cite{Abreu:2019yak, Abreu:2020ddv} and SND@LHC~\cite{Ahdida:2020evc, Ahdida:2750060}, will begin operation. Both experiments are placed in previously unused side tunnels, TI12 for FASER$\nu$ and TI18 for SND@LHC, which are located on opposite sides of the ATLAS experiment and about 500~m downstream from its collision point. At these locations, the experiments can be placed at or near the center of the neutrino beam. These detectors, which have a target mass of about 1~ton each, are expected to detect thousands of neutrino interactions during LHC Run 3.

FASER$\nu$ and SND@LHC also pave the way for a high-energy neutrino physics program during the high-luminosity LHC (HL-LHC) era. The FPF has been proposed as a facility to house a suite of experiments to fully explore the associated physics potential~\cite{Anchordoqui:2021ghd}. As shown in \cref{fig:Feynman}, it would be located about 620m downstream from the ATLAS interaction in a purpose-built cavern which surrounds the neutrino beam. Three neutrino experiments have been proposed for this facility: i) FASER$\nu$~2, which is a 20~ton emulsion-based neutrino detector, ii) Advanced SND@LHC, which is a 2-10 ton electronic neutrino detector, and iii) FLArE, which is a 10-100 ton liquid argon time projection chamber. In addition, the FPF includes FASER~2, a magnetized spectrometer to search for long-lived particles~\cite{Ariga:2018uku}, and FORMOSA, a plastic scintillator array to search for milli-charged particles~\cite{Foroughi-Abari:2020qar}. If placed behind the neutrino detectors, the spectrometer of FASER~2 could be used to identify the charges of muons exiting the neutrino detectors. 
\medskip 

As input for our study, we use the neutrino fluxes presented in Ref.~\cite{Batell:2021aja} for a $1~\m \times 1~\m$ cross sectional area considered for the FLArE-10 benchmark. For this, the collisions were simulated using the event generator \texttt{Sibyll~2.3d}~\cite{Riehn:2019jet} as implemented in \texttt{CRMC}~\cite{CRMC}, and the propagation and decay of long-lived hadrons is modelled by the fast neutrino flux simulation introduced in Ref.~\cite{Kling:2021gos}. \cref{fig:TotalRate} presents the total event rate in a 10 ton detector assuming an integrated luminosity of $3~\iab$ that is expected to be collected at the HL-LHC. We compare the event rate of events with final-state muons (blue) for $\lambda_{\mu\mu} = 1$ with the rate for final-state tau events (orange) for $\lambda_{\mu\tau} = 1$. Even for $m_\phi = 10$ GeV, more than $10\times \lambda_{\mu \beta}^2$ events are expected in each case.

In the next section, we will develop an analysis which aims to isolate the signal process, $\nu_\mu\, N \to \mu\, \phi\, X$, from the SM CC background, $\nu_\mu\, N \to \mu\, X$. To this end, we generate the BSM physics signal events using \texttt{MG5\_aMC}~\cite{Alwall:2014hca}, and the SM background events using \texttt{Pythia~8.2}~\cite{Sjostrand:2006za, Sjostrand:2014zea}. In the next step, we smear the final state momenta to approximate the effects of a finite detector resolution. In order to keep our analysis as independent of whatever detector(s) end up occupying the FPF, we make a minimal set of assumptions regarding their performance. We consider two simplified scenarios for the reconstruction capability of the FPF detector. In the optimistic scenario, we assume a hadronic energy resolution at the $15\%$ level as an optimistic target, whereas in the less optimistic scenario the resolution is 45\%. In both cases, we assume the muon energy resolution of the detector is at the $5\%$ level. We use these as smearing parameters for the final state particles in the MC simulation while keeping the direction of three momentum intact. For simplicity, we further assume a perfect acceptance and lepton identification rate.

\section{Muon Final-State Analysis}\label{sec:AnalysisMuon}

\begin{figure*}
    \centering
    \includegraphics[width=0.98\linewidth]{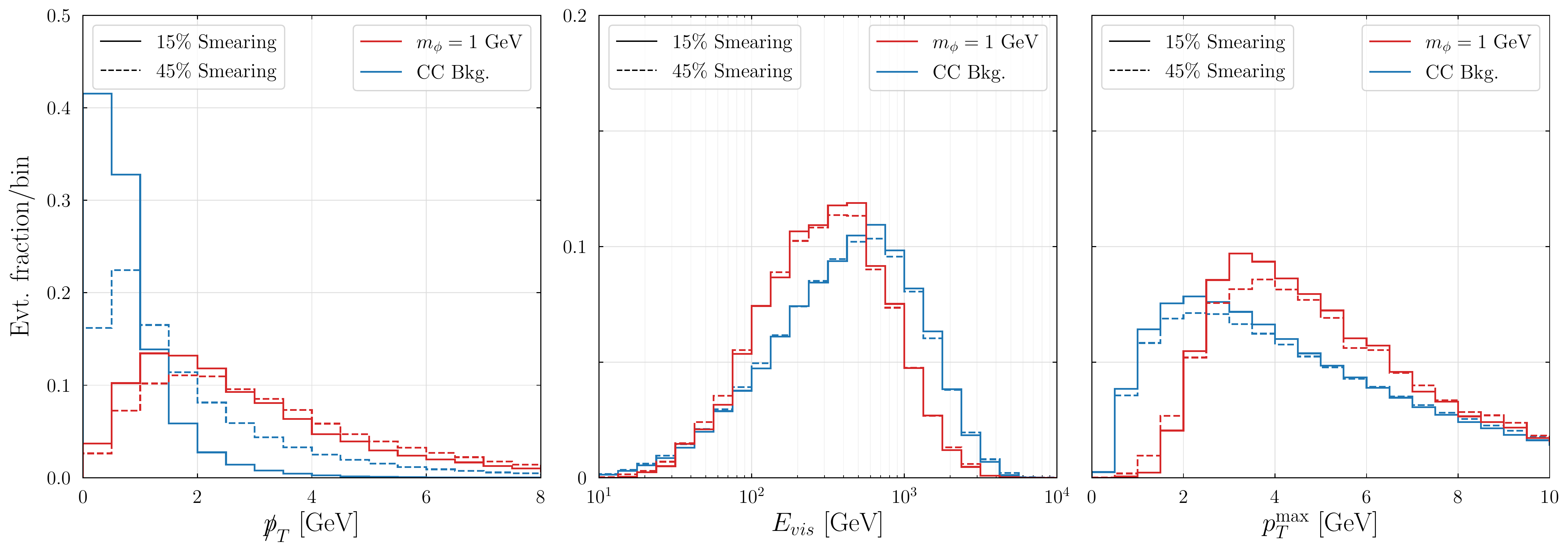} \\
    \caption{One-dimensional distributions of signal (red histograms for $m_\phi = 1$ GeV) and background (blue histograms) event rates as a function of the missing transverse energy (left), the visible energy (center), and the largest transverse momentum of visible particles (right). In each panel, the solid (dashed) distributions correspond to 15\% (45\%) smearing of the final state quark momenta, and the distributions are normalized such that the total bin count sums to 1. Note the logarithmic axes in the center panel.\label{fig:SigBkgDists}}
\end{figure*}

In this section, we present the approach for analyzing events where the final states are characterized by a muon along with large missing transverse momentum (MET), and the resulting sensitivity to the neutrinophilic coupling $\lambda_{\mu\mu}$ as a function of the  scalar mass $m_\phi$ with the detector capabilities assumed above.
As discussed in \cref{sec:Simulation}, the signal consists of the process $\nu_\mu\, N \to \mu^+\, \phi\, X$, and the background is dominated by the regular charged-current process $\nu_\mu\, N \to \mu^-\, X$.
In the analysis, we also take into account the charge-conjugated processes with incident anti-neutrinos. 
We assume that the final state muon can be positively identified by the detector, as long as they are energetic enough to travel further than several hadronic interaction lengths (to reject the possibility of it being a charged pion). This amounts to requiring $E_\mu \gtrsim 1\, {\rm GeV}$ for selected events~\cite{PDGpenetration,PDGpenetration2}. This allows for rejection of backgrounds from neutral-current neutrino scattering, among other types.

To appreciate MET as a useful signature for discriminating between signal and background events, we first present the differential cross section of the mono-neutrino scattering process at the parton level,
\begin{equation}\label{eq:dsigma}
    \begin{split}
\frac{d\sigma_{\nu_\mu u \to \phi \mu^+ d}}{d p_{T_\phi}} &\simeq 
\frac{3|\lambda|^2 G_F^2 s p_{T_\phi}^3}{8\pi^3 m_\phi^4} \\
&\hspace{-0.5cm}\times\left[ \left( 1 +  \frac{2p_{T_\phi}^2}{m_\phi^2}\right) \log \left( 1 +  \frac{m_\phi^2}{p_{T_\phi}^2}\right)- 2 \right]  ,
\end{split}
\end{equation}
where $p_{T_\phi}$ is the transverse momentum of the outgoing $\phi$ particle with respect to the incoming neutrino beam. We have made the approximation $p_{T_\phi}, \ m\ll \sqrt{s}$, where $m$ represents all the mass parameters involved in this process. A  nonzero $\phi$ mass is kept here to regularize the collinear divergence of $\phi$ radiation in the limit $p_{T_\phi}\to0$.
The differential cross section peaks around $p_{T_\phi} \simeq m_\phi$. In the region $p_{T_\phi} \gg m_\phi$ the differential cross section goes as 
$1/p_{T_\phi}$.
After production, $\phi$ subsequently decays into neutrinos or DM and appears invisible. The resulting missing transverse momentum is $\cancel{p}_T = p_{T_\phi}$, whose distribution is identical that of $p_{T_\phi}$. Our MC simulation results confirm the behavior of Eq.~\eqref{eq:dsigma}

After passing the MC simulated event samples through the detector smearing, as detailed in the previous section, we introduce the following set of kinematic observables for signal-background comparison,
\begin{itemize}
    \item $\slashed{p}_T$, the missing transverse momentum, reconstructed from visible final state particles,
    \item $E_{\rm vis}$, the total energy of all visible particles,
    \item $p_T^\text{max}$, the highest transverse momentum of visible final state objects.
\end{itemize}

\begin{figure*}[t]
    \centering
     \includegraphics[width=0.85\linewidth]{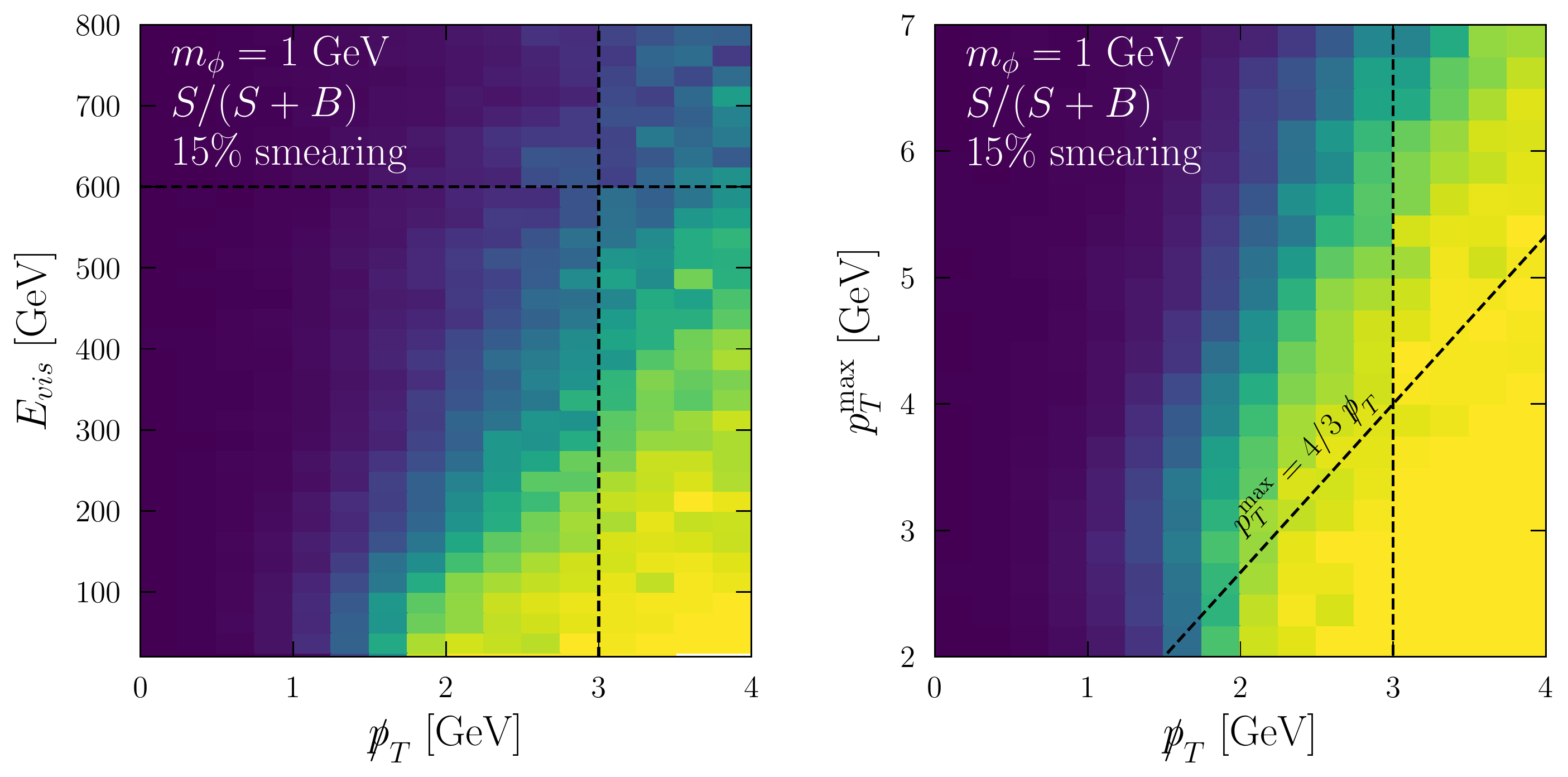}
     \includegraphics[width=0.85\linewidth]{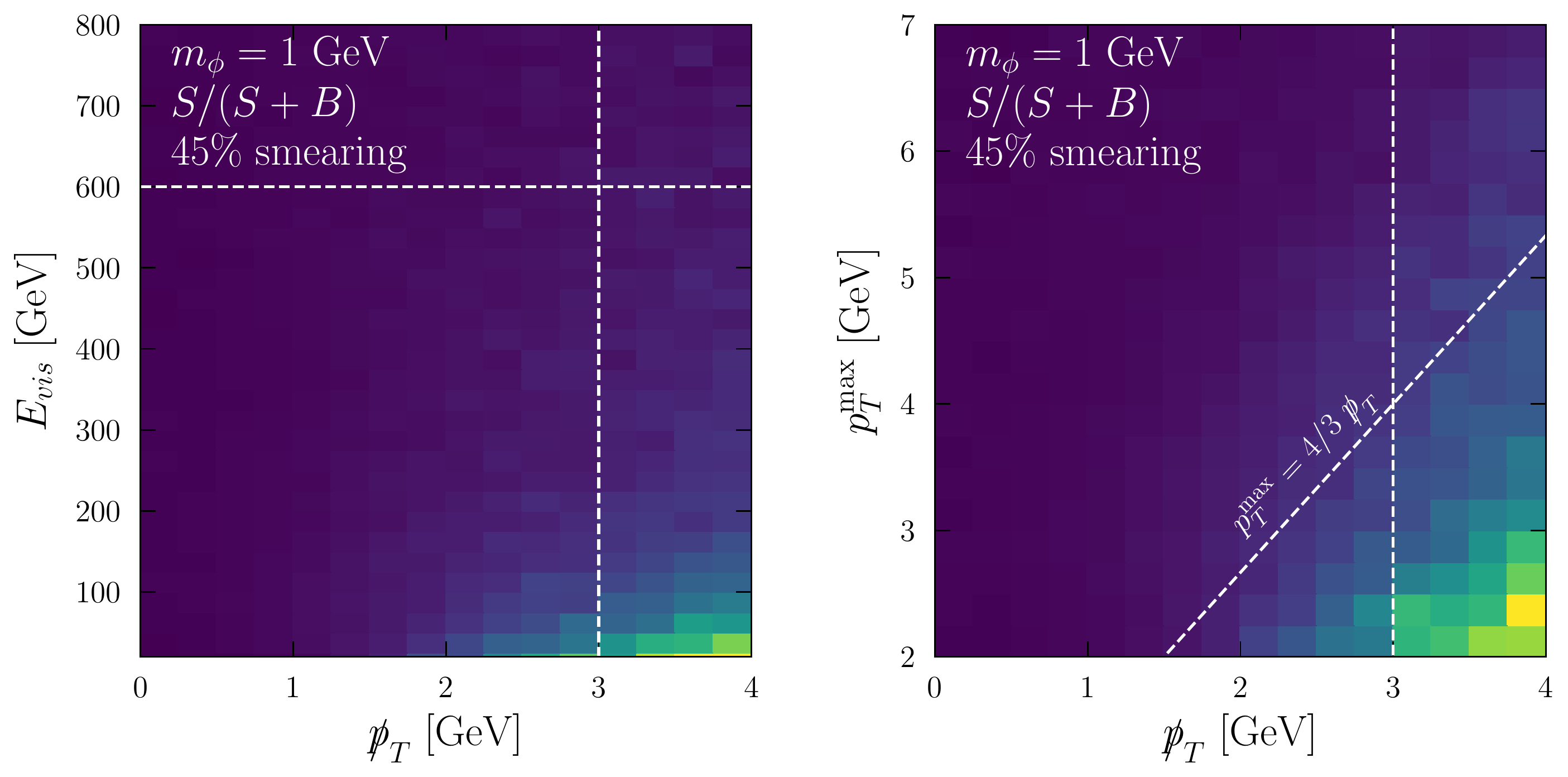}
    \caption{Signal to signal-plus-background ratio ($S/(S+B)$) in the $\slashed{p}_T - E_{\rm vis}$ (left) and $\slashed{p}_T - p_T^\text{max}$ (right)  planes, comparing a $m_\phi = 1$ GeV signal with the charged-current background. The upper and lower panels correspond to a detector with hadronic energy resolution of 15\% and 45\%, respectively. Yellow (purple) regions correspond to where $S/(S+B)$ is larger (smaller). \label{fig:SigBkgRatio}}
\end{figure*}

\cref{fig:SigBkgDists} displays one-dimensional distributions of
signal (red, for $m_\phi = 1$ GeV) and background (blue)
as a function of $\slashed{p}_T$ (left), $E_{\rm vis}$ (center), and $p_T^\text{max}$ (right). In each panel, all distributions are normalized such that the bin counts sum to 1.
We see here, as first noticed in Refs.~\cite{Berryman:2018ogk,Kelly:2019wow}, that relative to background, the signal is more pronounced for large $\slashed{p}_T$ and $p_T^\text{max}$, but small $E_{\rm vis}$. 
To better take advantage of the correlation between multiple observables, \cref{fig:SigBkgRatio} shows two-dimensional distributions,
the ratio between signal to signal-plus-background events, $S/(S+B)$,
in the $\slashed{p}_T - p_T^\text{max}$ and $\slashed{p}_T - E_{\rm vis}$ planes, for both 15\% and 45\% 
hadronic energy resolution scenarios.
The ratio is higher (lower) for yellow (purple) regions in the figure. For both \cref{fig:SigBkgDists} and \cref{fig:SigBkgRatio}, we use arbitrary scaling/normalization so that the results hold for any $\lambda_{\mu\mu}$. 

To demonstrate the power of these variables in discriminating between the charged-current backgrounds and our proposed signal, Table~\ref{tab:CutandCount} presents the results of a simple cut-and-count based analysis using $m_\phi = 1$ GeV as a benchmark point. The efficiency factors quoted in the table correspond to 15\% (45\%) smearing in hadronic energy resolution.
For the 15\% smearing case, we find these cuts very powerful and can reduce the charged-current background by five orders of magnitude, while retaining more than 10\% signal efficiency.
After a cut on $\slashed{p}_T$ the background rate is already reduced by three orders of magnitude, showing that MET is a powerful observable for probing the neutrinophilic scalar.

{\renewcommand{\arraystretch}{1.5} 
\begin{table}
    \centering
    \begin{tabular}{c||c|c|c}
    \hline
    \hline
    & $E_{\rm vis.} < 600$ GeV & $\slashed{p}_T > 3$ GeV & $p_T^\text{max} < \frac{4}{3}\slashed{p}_T$ \\ \hline \hline
    $\nu_\mu + \overline{\nu}_\mu$ CC & 61\% (62\%) & 0.2\% (6\%)& $10^{-5}$ (1\%)\\ \hline
    $m_\phi = 1$ GeV & 76\% (76\%) & 26\% (33\%) & 15\% (18\%)\\
    \hline
    \hline
    \end{tabular}
    \caption{Fraction of surviving background and signal ($m_\phi = 1$ GeV) events undergoing subsequent cuts on the visible energy, missing transverse energy, and largest transverse momentum of a visible particle, for 15\% (45\%) smearing on the quark momenta.}
    \label{tab:CutandCount}
\end{table}}

\begin{figure*}[t]
\centering
\includegraphics[width=0.485\textwidth]{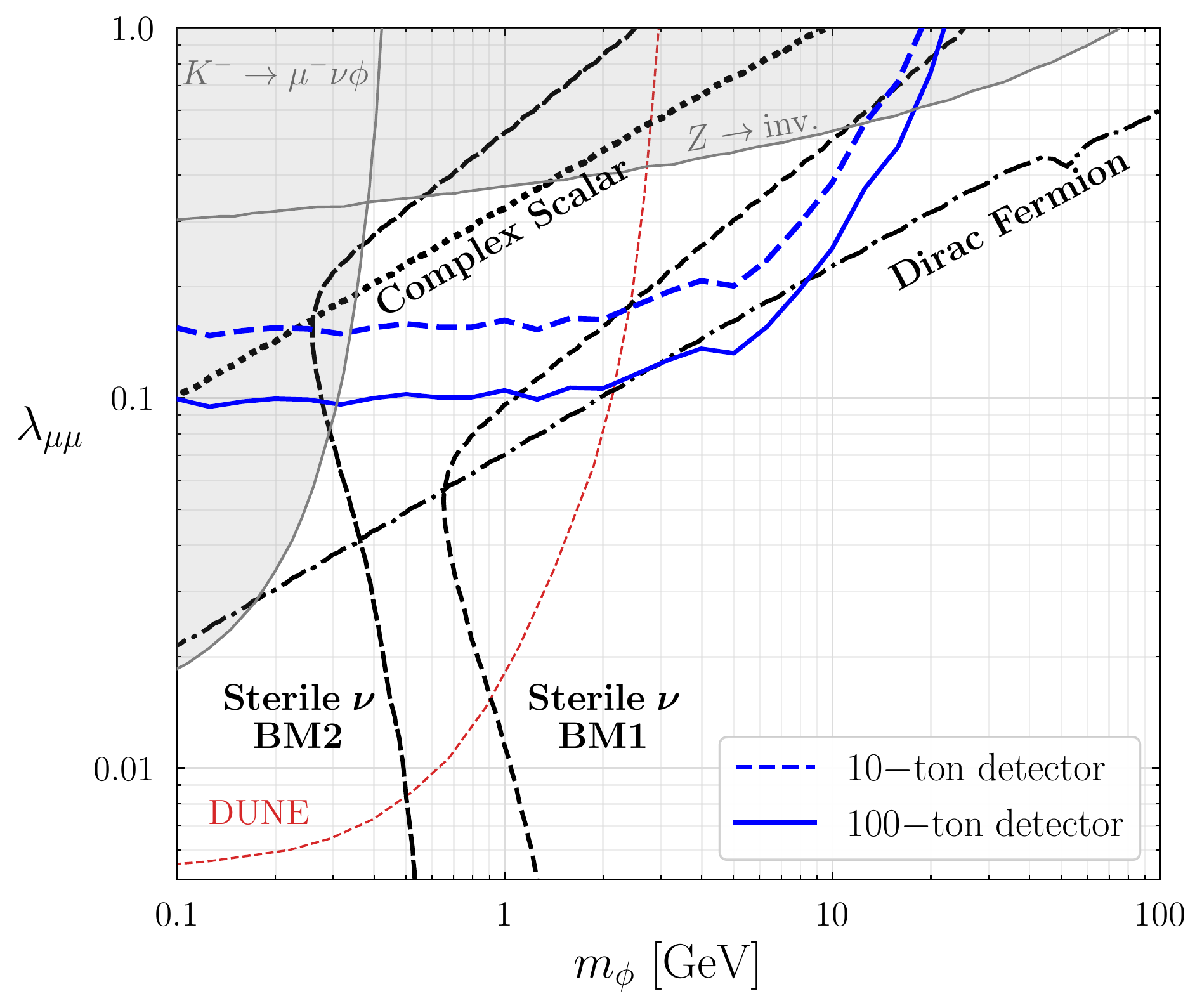}~
\includegraphics[width=0.485\textwidth]{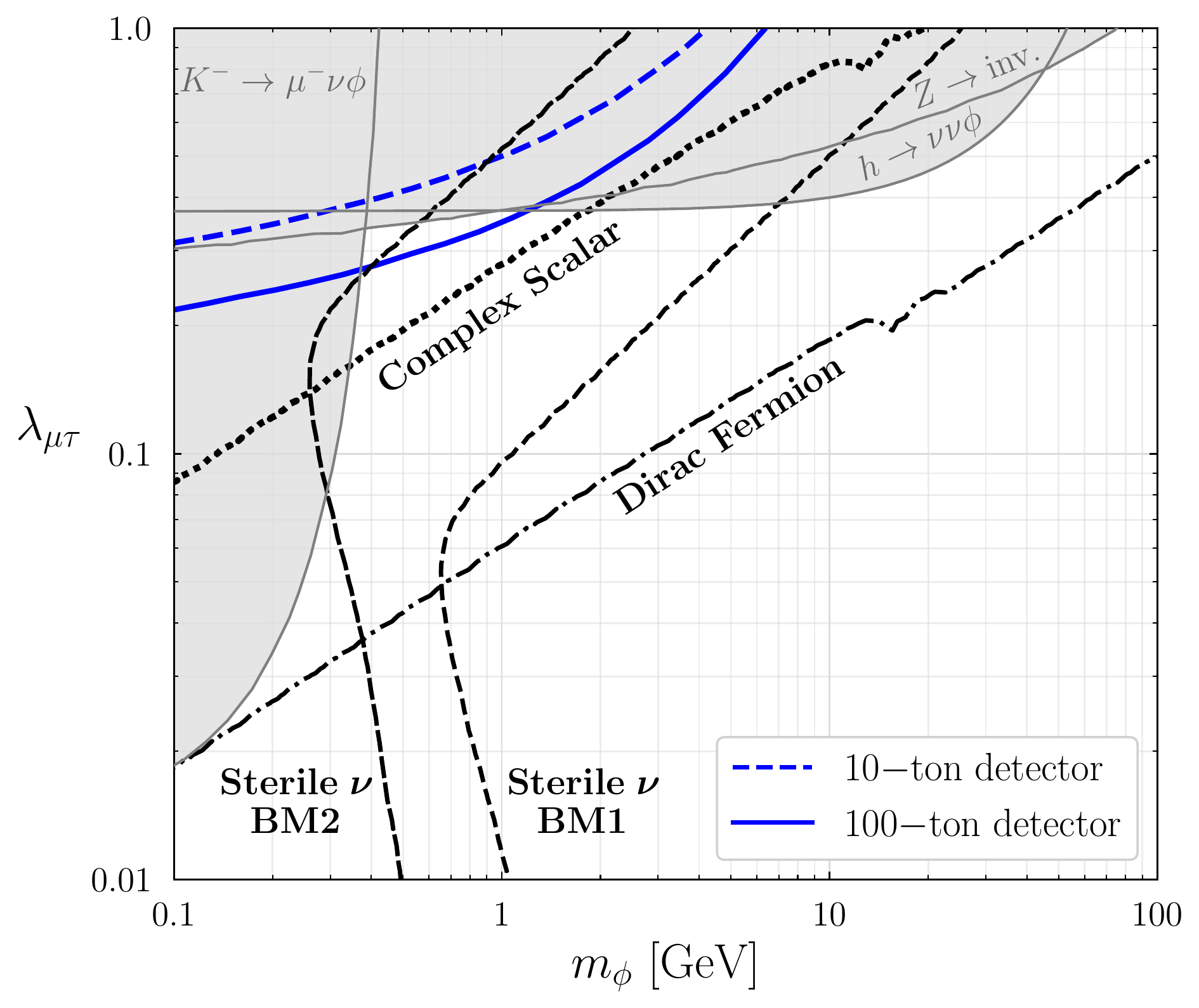}
\caption{Expected reach of a FLArE-like detector using the mono-neutrino signature with final-state muons (left) and taus (right). The dashed (solid) blue curves correspond to a 10 (100) ton detector with a 1m x 1m cross sectional area and $3\,{\rm ab}^{-1}$ of $pp$ collisions at High-Luminosity LHC. The gray shaded regions are existing constraints from invisible decays of kaons, the $Z$ boson, and the Higgs boson, while the dashed red curve in the left plot is the expected reach of DUNE. 
}\label{fig:Bounds}
\end{figure*}

To optimize our results, we feed these three observables into a neural network and determine a cut on the resulting $S/\sqrt{B}$ ratio to maximize sensitivity to the $\lambda_{\mu\mu}$ coupling \cite{keras}.
Our main results are presented in \cref{fig:Bounds}, where the left plot shows the expected sensitivity in $\lambda_{\mu\mu}$, assuming a detector mass of 10 ton (dashed blue curve) and 100 ton (solid blue curve), and 15\% energy smearing on the outgoing hadronic final state. This sensitivity to $\lambda_{\mu\mu}$ is unmatched by the existing charged-meson decay constraints (gray shaded region) above $m_\phi \approx 250$ MeV, and surpasses the expected DUNE sensitivity (red dashed curve) for $m_\phi \gtrsim 2$ GeV. The sensitivity also exceeds existing constraints from the invisible widths of the $Z$ and Higgs bosons (gray shaded region) for $m_\phi$ up to $\sim 20$ GeV. 

In the same figure, we also show the dark matter targets as discussed in \cref{sec:ModelDM} with black curves of various styles. In particular, the thermal freeze out targets are shown by the black dot-dashed and dotted curves where the dark matter is a Dirac fermion and complex scalar, respectively, defined in \cref{subsec:thermalDM}. On these curves, the 
Planck-observed value of dark matter relic density~ \cite{Aghanim:2018eyx} is successfully produced.
Here, we fix the additional parametric dependence using 
\begin{equation}
    m_\phi = 3 m_\chi, \quad y = \lambda_{\alpha\beta} \ .
\end{equation}
For higher dark matter-mediator coupling $y$, both curves will shift to smaller $\lambda_{\mu\mu}$ values. However, for the case of complex scalar with $Z_3$ symmetry, we find that even for $y \sim \mathcal{O}(1)$ the target curve still lies within the reach of the FPF detector thanks to the extra phase space suppression of the $2\to3$ annihilation. 

On the other hand, the sterile neutrino dark matter targets, whose relic density is driven by neutrino self-interaction, are shown by the black dashed curves. Here we consider two benchmark values for dark matter mass $m_4$ and active-sterile mixing $\theta$,
\begin{itemize}
    \item BM1: $m_4 = 7.1$ keV, $\sin^2(2\theta) = 7 \times 10^{-11}$ 
    \item BM2: $m_4 = 16$ keV,\; $\sin^2(2\theta) = 8\times10^{-14} $ 
\end{itemize}
Their unusual shape is due to a close interplay between $m_\phi$ and temperature dependence in $\Gamma$ and $\theta_{\rm eff}$~\cite{deGouvea:2019phk,Kelly:2020aks,Kelly:2020pcy}. It is worth pointing out that for each S$\nu$DM mass we have chosen the active-sterile mixing angle to be close to its maximally allowed value by indirect X-ray searches. As a result, there is not much room left for these relic curves to move to the smaller $\lambda_{\mu\mu}$ direction. The FPF detector can serve as a very powerful tool at probing such a DM production mechanism. 

Our results show that future FPF detectors have the exciting potential to investigate both types of dark matter targets that are unconstrained by existing experiments.
In the thermal freeze out case where mediator $\phi$ with equal coupling to neutrino and complex scalar dark matter, the whole relic curve can be covered. 
The Dirac fermion dark matter target is out of reach of the 10-ton detector but part of it is within reach of a 100-ton one.
For S$\nu$DM freeze in, the FPF detector will be able to cover the BM1 curve above $m_\phi \approx 1-2$ GeV, which is complementary to the future DUNE experiment which can cover the lower $\phi$ mass regions.

\begin{figure}[t]
\centering
\includegraphics[width=0.45\textwidth]{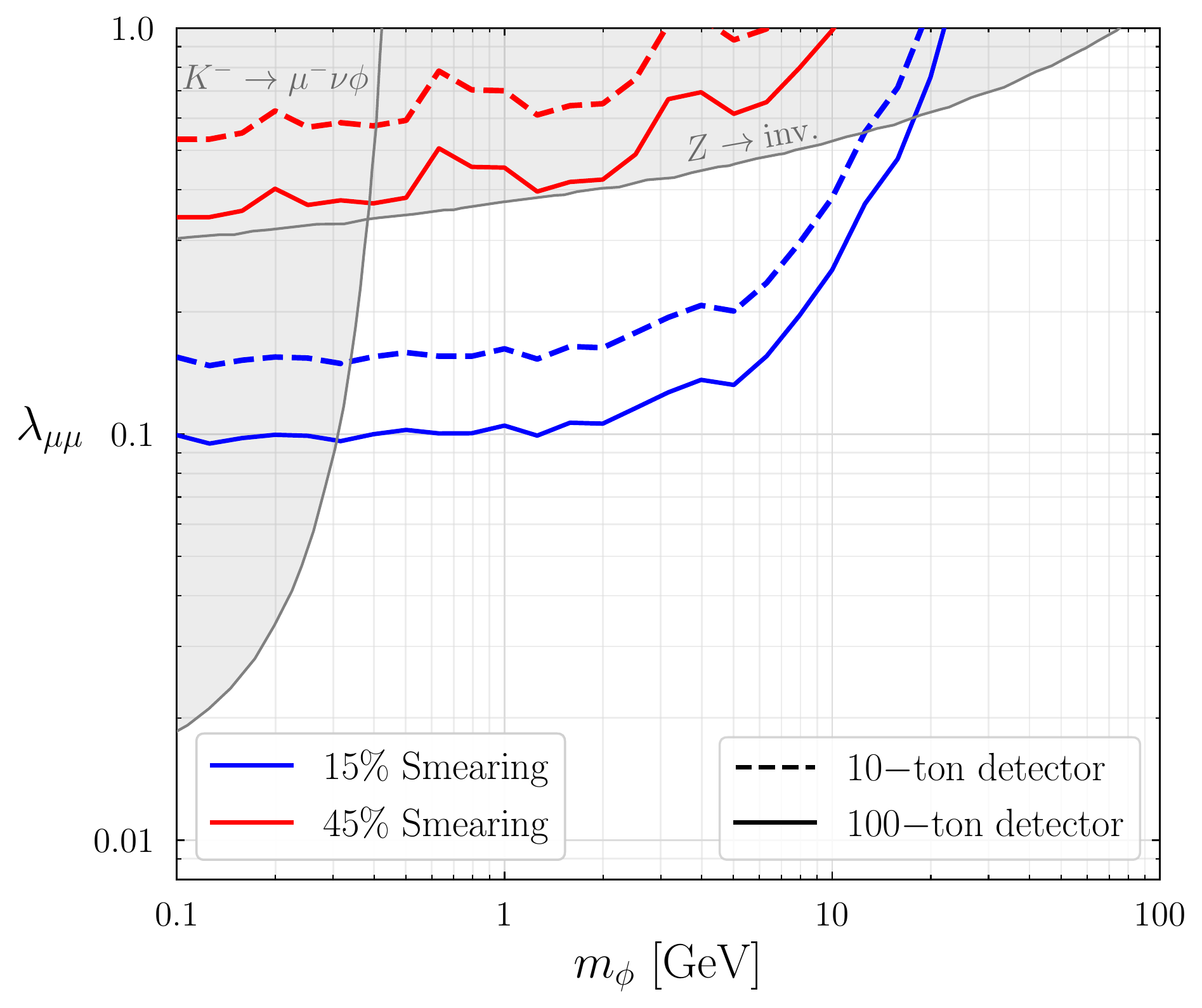}
\caption{Expected reach of a FLArE-like detector using the mono-neutrino signature with final-state muons assuming 15\% (blue curves) and 45\% (red curves) smearing on the final state hadron momenta. The dashed (solid) curves correspond to a 10 (100) ton detector. The gray shaded region are existing constraints from invisible decays of kaons, the $Z$ boson, and the Higgs boson. 
}\label{fig:Bounds_45}
\end{figure}

In presenting \cref{fig:Bounds}, we have focused on our optimistic scenario of energy resolution of 15\%. 
For the less optimistic case with a 45\% energy resolution, we find the sensitivities weaken substantially, as depicted by the red curves in \cref{fig:Bounds_45}, which can no longer exceed existing constraints.
This comparison quantifies the requirement for future FPF detectors to be able to probe the neutrinophilic interaction using MET as the key discriminator between signal and background. 

We have also examined the sensitivity of the existing FASER$\nu$ detector and reached a similar conclusion due to its limited detector mass and energy resolution. However, we realize that our current imagination of future  detectors in the FPF hall is limited. 
The bottom line is that our analysis above and \cref{fig:Bounds} show that there is vast parameter space separating the current limits and the potential future sensitivity of FPF detectors and awaiting upcoming experimental breakthroughs.

\section {Tau Final-State Analysis}\label{sec:AnalysisTau}

We now turn to the scenario where only the off-diagonal coupling $\lambda_{\mu\tau}$ is non-zero. The signal process for this case is $\nu_\mu\, N \to \tau^+\, \phi\, X $, whereas the dominant background being the charged-current process $\nu_\tau\, N \to \tau^-\, X$. The $\lambda_{\mu\tau}$ could also lead to muon + MET signal with an incoming $\nu_\tau$. We will not consider this possibility because the flux of $\nu_\tau$ is more than three orders of magnitude lower than the flux of $\nu_\mu$, leading to much smaller signal rates. The same argument also helps to suppress the above background for $\tau$ lepton production. With a 10-ton detector at FPF, the expected number of background $\tau$ is $\mathcal{O}(1000)$. An excessive $\tau$ appearance over the SM expectation would then be a sign of new physics. 

In practice, however, there are sizable  uncertainties, associated with the modeling of the neutrino flux, neutrino interaction, and tau identification capabilities, which affect the search for tau neutrinos. In particular, current predictions include a large systematic uncertainty on the normalization of the tau neutrino flux~\cite{Bai:2020ukz, Kling:2021gos}. Nevertheless, the spatial distribution of this signal compared to the SM background can be a useful handle to reduce these systematic uncertainties. The SM tau neutrino background is expected to have a broader spatial distribution, as it arises from $D_s$ meson decay leading to a typical transverse momentum of $p_T \sim m_D$. In contrast, the $\nu_\mu$ flux that induces our signal originates from from lighter meson decays. The transverse momenta of muon neutrinos are therefore typically smaller, $p_T \sim m_\pi/m_K$, leading to a more narrow beam. 

We use this feature when performing the analysis, and define two samples: i) a signal region covering the inner $50~\cm \times 50~\cm$ cross sectional area around the center of the beam, and ii) a control region covering the remaining of the $1~\m \times 1~\m$ cross sectional area around the signal region. We assume that the control region, which is largely dominated by the SM CC tau neutrino background, is used to constrain the background normalization. We can then use the signal region to search for an excess of events with a $\tau$ in the final states. To estimate the sensitivity to the coupling $\lambda_{\mu\tau}$ as a function of $m_\phi$, we then count the number of such events assuming that the uncertainty in the background event rate is statistics-dominated. The resulting sensitivity curves are shown in the right panel of \cref{fig:Bounds}, assuming a 10 ton (dashed blue curve) or 100 ton (solid blue curve) detector. We can see that the sensitivity of 10 ton detector is slightly weaker than the existing constraints (the latter are similar to the $\lambda_{\mu\mu}$ case), but a 100 ton detector will be able to start exploring new parameter spaces and test the sterile neutrino dark matter target (BM2). 

\section{Discussion \& Conclusions}\label{sec:Conclusions}
As our precision understanding of neutrinos develops into the coming decades, it is imperative to test all SM and beyond-the-SM properties that they have, or may have. Neutrino self-interactions sourced by a new mediator are one such testable property, and the repercussions in the event of a discovery of these would be momentous.
With the future Forward Physics Facility in development, a new era of high-energy, laboratory-based neutrino physics is on the horizon. The detectors planned for this facility are wide in scope and offer a variety of physics opportunities in studying both Standard Model and beyond-the-Standard-Model physics. One particular advantage of the FPF is the unparalleled energy scale of the neutrinos being studied compared to other laboratory-based environments. We have demonstrated that this high-energy flux will allow for searches for new, neutrinophilic force-carriers that could serve as a portal between the SM and DM.

Such new force-carriers can connect the neutrinos to a variety of well-motivated DM model realizations, including both thermal freeze-out and sterile-neutrino dark matter freeze-in mechanisms. We have considered several benchmark models and demonstrated that the FPF, collecting data coincident with $3$ ab$^{-1}$ of LHC luminosity and a 10-ton or 100-ton liquid argon detector, has excellent sensitivity to probe these models. Key in performing these searches is the ability to measure the hadronic energy in neutrino events at the ${\sim}15\%$ level. Comparing against existing and upcoming experimental constraints, the FPF can search for neutrinophilic mediators with higher mass (up to ${\sim}$20 GeV) with couplings of order $\mathcal{O}(0.1)$ to the active neutrinos. This pushes searches to heavier masses relative to those that can be performed at DUNE, and to smaller couplings than those excluded by studying $Z$ boson decays. The findings in this work present a nice complementarity among these frontiers.

\acknowledgements

We are grateful to the authors and maintainers of many open-source software packages, including 
\texttt{hepmcio}~\cite{buckley_andy_2018_1304136}, 
\texttt{Keras}~\cite{chollet2015keras},
\texttt{pylhe}~\cite{lukas_2018_1217032}, 
and \texttt{scikit-hep}~\cite{Rodrigues:2019nct}. The work of FK is supported by the U.S.~Department of Energy under Grant No.~DE-AC02-76SF00515 and by the Deutsche Forschungsgemeinschaft under Germany’s Excellence Strategy - EXC 2121 Quantum Universe - 390833306. Fermilab is operated by the Fermi Research Alliance, LLC under contract No. DE-AC02-07CH11359 with the United States Department of Energy. DT and YZ are supported by the Arthur B. McDonald Canadian Astroparticle Physics Research Institute.


\bibliography{references}

\end{document}